%% file: main.tex
\documentclass[sigconf]{acmart}
\usepackage{prelim2e}

\input{header}

\title{\textsc{SchemaDB}: Structures in Relational Datasets}

\makeatletter
\newcommand{\linebreakand}{%
  \end{@IEEEauthorhalign}
  \hfill\mbox{}\par
  \mbox{}\hfill\begin{@IEEEauthorhalign}
}
\makeatother

\author{Cody James Christopher}
\orcid{0000-0001-8444-2292}
\affiliation{%
    \institution{CSIRO Data61}
    \city{Canberra}
    \state{ACT}
    \country{Australia}
    }
\email{cody.christopher@data61.csiro.au}
\affiliation{Cyber Security Cooperative Research Centre}

\author{Kristen Moore}
\orcid{0000-0002-9962-5080}
\affiliation{%
    \institution{CSIRO Data61}
    \city{Melbourne}
    \state{VIC}
    \country{Australia}
    }
\email{kristen.moore@data61.csiro.au}
\affiliation{Cyber Security Cooperative Research Centre}

\author{David Liebowitz}
\affiliation{%
\institution{Penten Pty Ltd}
    \city{Canberra}
    \state{ACT}
    \country{Australia}
    }
\email{david.liebowitz@penten.com}
\orcid{0-0-0-0}
\affiliation{Cyber Security Cooperative Research Centre}

\begin{document}
\begin{abstract}
In this paper we introduce the \textsc{SchemaDB} data-set; a collection of relational database schemata in both \texttt{sql} and graph formats. Databases are not commonly shared publicly for reasons of privacy and security, so schemata are not available for study. Consequently, an understanding of database structures in the wild is lacking, and most examples found publicly belong to common development frameworks or are derived from textbooks or engine benchmark designs. \textsc{SchemaDB} contains 2,500 samples of relational schema found in public repositories which we have standardised to \texttt{MySQL} syntax. We provide our gathering and transformation methodology, summary statistics, and structural analysis, and discuss potential downstream research tasks in several domains.
\end{abstract}

\maketitle
\begin{acks}
The work has been supported by the Cyber Security Research Centre Limited whose activities are partially funded by the Australian Government’s Cooperative Research Centres Programme.
\end{acks}

\input{10_intro}

\input{30_curation}
\input{40_analytics}
\input{50_application}

\section{Conclusion}
Database schema data sets are needed for various ML applications, including to automate and scale the synthesis of databases for use in cyber deception. \textsc{SchemaDB} is intended to enable such research, as well as to provide a standardised example for other potential data set providers. A clear limitation of \textsc{SchemaDB} is that it is restricted to freely available, public datasets, and the types of schemas accessible on GitHub may not be representative of all segments of the database population existing on the internet, within large corporations, and proprietary commercial applications. We therefore hope that the release of \textsc{SchemaDB} will encourage others to release similar datasets to augment our initial release. There is also the potential to grow SchemaDB by rerunning the data collection and ETL scripts periodically over time, and also by incorporating other code repository hosts like GitLab. 

\citestyle{acmnumeric}
\bibliographystyle{ACM-Reference-Format}
\bibliography{main}
\end{document}

%% file: header.tex
\usepackage[utf8]{inputenc}
\usepackage{svg}
\usepackage{listings}

%% file: 10_intro.tex
\section{Introduction}
An on-going problem in the machine learning research community is a shortage of suitable datasets. Often, the release of data is impeded by concerns surrounding intellectual property, disclosure, or privacy. This hampers efforts to replicate, extend, and compare results across a research domain.


One such example where publicly available data is lacking is that of database schemas. The majority of databases available publicly are designed for benchmark evaluation of the performance of various SQL engines, or developed for education purposes to demonstrate design principles. Unfortunately, these are not suitable representations from which to learn the structure of real world databases, such as those one might expect to find inside a corporate network. Other well-known dataset corpora such as Kaggle are a fantastic resource for large datasets usually designed for machine learning tasks, but these are typically not relational in nature. Those that could be are still usually presented flat in a machine learning digestible form.

A rich and standardised schema dataset would be useful for the study of common database structures, and would also provide the potential to train ML algorithms for database generation and simulation. Such generative ML models are of particular value in applications such as cyber deception and cyber range exercises, where bringing automation and scale to the generation of database schemas would save a large amount of time in hand curating realistic content.

To this end, we introduce \textsc{SchemaDB}, a collection of 2,500 standardised schemas collected from real projects, largely found on GitHub. We discuss our gathering and transformation methodology, and provide an analysis of the structure and summary statistics of the collection. We release the curated dataset (and extensions) alongside the paper, as well as the code used to extract schema from a collection of repositories.



\subsection{Existing Datasets}
To the best of our knowledge there is only one other public dataset that has similar purpose and viable schema -- the CTU Prague Relational Learning Repository~\cite{motl2015ctu}. The stated purpose of this dataset is to enable machine learning on relational data, as opposed to single table (the majority of data releases). It includes full databases (inclusive of data) on a public database server for this purpose. After excluding benchmarking or sample databases, there are 62 samples. We  incorporate these into \textsc{SchemaDB} as the only samples not from GitHub. The primary advantage of the CTU dataset is that full data is also available to facilitate machine learning tasks.

%% file: 30_curation.tex
\section{Dataset Curation}
This section will detail the process of collecting and curating \textsc{SchemaDB}.
\subsection{Collection \& Filtration}
The primary data source for \textsc{SchemaDB} is a collection of open source repositories found on GitHub. Sources were selected by searching for repositories containing an indicative \verb!schema.sql! file. Often these repositories contain a multitude of \verb|.sql| files, and regularly multiple \verb|schema| files. These files can serve a variety of purposes; targeting various SQL dialects, performing upgrade migrations, editing security and permissions, or (less often) performing data insertion. 

To avoid excessive manual filtration of thousands of files within thousands of repositories, only the most populous schema from each repository was captured as a representative sample, assuming that a larger schema is more likely to have some degree of normalisation applied for ease of management. The population of a schema here is determined by the number of tables present in the file. This does have the effect of introducing a few edge case effects where in certain cases migration files are captured when particularly large upgrades occur, but pre-selection statistics (number of valid files per repository), indicates that these are a small minority and can still represent valid sub-schema.

We consider only those schemas that can be parsed under standard encoding (\verb|utf8| and \verb|utf16|) as this covers the vast majority of conforming schema, although in future it may be worth extending this to capture schema in less common encodings as to not exclude those with names written in less common languages (where supported). As the primary purpose of this collection is to provide a basis for investigating relational structures, we filter initially to those schema that contain at least two tables such that relations can actually be defined. Edge cases were found in the raw data when commented code was detected indicating the presence of user intervention (as opposed to database engine generation as the result of a dump, or tool assistance). This is subsequently avoided by performing a secondary parsing of every file that passes the prior conditions.

\subsection{Graph Transform \& Canonisation}
To standardise the data format for subsequent analysis, we modified the existing grammars and constructed custom parsers for three major SQL dialects; \verb|MySQL|, \verb|SQLite|, and \verb|postgreSQL|. As the most permissive of the dialects by design, \verb|MySQL| was selected as the output format for canonisation. To assist with graph analytics, a directed heterograph representation was generated from the initial parse, where necessary alterations could be performed such that valid SQL could be generated in the target dialect. This included performing a data type mapping as part of the transpilation. 

\begin{figure}[tbp]
    \centering
    \includegraphics[scale=0.5]{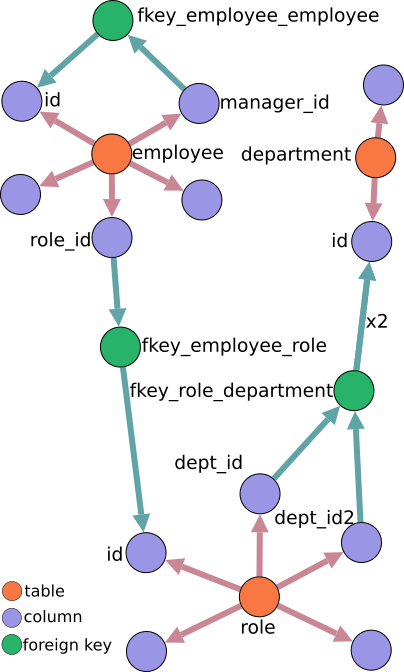}
    \caption{Graph representation of a relational schema. Some nodes left unlabelled for brevity}
    \label{fig::empgraph}
\end{figure}

Graphs are constructed by creating nodes for each table, column, and foreign key linkage, where nodes are explicitly typed. Edges in these graphs are not explicitly typed but can be inferred from the direction and the connecting nodes. Edges outbound from a table node to column node indicate that the column belongs to the table. Edges outbound from a column to a foreign key node indicate that this is a \emph{referencing} column, and edges outbound from a foreign key to a column indicate a \emph{referenced} column. Foreign key nodes can have multiple inbound or outbound edges in the case of compound foreign keys (involving multiple columns), but as a collection the columns on either side will always be incident to the same source and destination tables, and the inbound and outbound degrees will be identical. Nodes also contain the semantic information associated with their type. For tables this information includes the table name and primary key(s), for columns this is their name, data type, optional data length, and position, and for foreign keys this is the name of the constraint (if explicit) and ordered lists of the columns in the key as this cannot be derived from the edges. We illustrate this graph views in a simple schema in Figure~\ref{fig::empgraph}.

For a higher level dependency analysis, what we denote a \emph{skeleton} of this graph is provided. The skeleton is a regular graph, where nodes denote tables and directed edges denote foreign key dependencies. The edge direction here is reversed, with an outbound edge indicating that the destination \emph{references} the source. 

Additional parsing rules were utilised as many dialects support declaring key constraints separately to table creation, and some methods of automated schema dumps perform this automatically. Some dialects also often support designing multiple sub-schemas in the same database. For ease of analysis, we count these as a single database entities and adjust duplicate table names as necessary. Some manual filtration of outliers at the extreme ends of the distribution (large amount of tables, or single tables) was carried out to remove schemas designed for benchmarking, and these are not present in the released data.

At this stage each graph skeleton is then topologically sorted and this order is subsequently used for canonisation to valid \verb|MySQL|, with obfuscation of the database name such that individual source repositories are not easily identifiable. Topologically sorting ensures that the tables are created in an order conducive to foreign key enforcement (the referenced column exists prior to the referencing column), although this is not strictly necessary when the entire database is created with a single transaction.

For a final consistency check, we make use of existing techniques to check for duplication amongst the corpus. Existing software approaches designed for plagiarism analysis are suitable here as they support detection of structure even in the presence of changing variable names. Specifically, we make use of the Measure of Software Similarity (MOSS)~\cite{schleimer2003winnowing} as the primary duplication detection method. When a high degree of similarity is detected, only one sample amongst all identified duplicates is kept for the canonical set. However we provide the degree of duplication and the excluded samples separately, should they be deemed necessary. Roughly 30 samples were removed as a result of this


\subsection{Heuristic Augmentation}
After an initial canonisation pass described in the previous section, further analysis and curation is possible. For example, during analysis it was discovered that a not insignificant number of samples containing more than two tables were devoid of any sort of foreign key. Whilst it would be ideal to perform a normalisation analysis, there is presently no way to ascertain this perfectly without performing an analysis of functional dependencies on every sample manually, which we consider a downstream task. It is, however, likely that samples in the corpus are not normalised according to common practice.

\subsection*{Missing Foreign Key Imputation}
One avenue of augmentation is to try and find implied (missing) foreign keys. As a result of common naming conventions, we demonstrate that it is possible to make explicit these implicit keys. Consider that common practice is to have an \verb|ID| column as the primary key of entity, which we denote \verb|entityA| for the purposes of the example. Should there be another table which contains references to this entity, the keyed column will typically directly reference this primary key and subsequently the column name will reflect this: \verb|entityA_ID| or \verb|entityAID| or similar.

We show the results of performing this type of analysis under a variety of heuristics. Firstly, we consider only direct matches -- where a column name is exactly the underscore concatenation of another table name with a column in that table. We then allow a progressive increasing of the standard edit (Levenshtein) distance between these generated identifiers to take account of common variations, acronyms and idioms. We observe than beyond a distance as small as three, the frequency of spurious matches becomes untenable due to prevalence of short column names and abbreviations. We provide in the release the exact (or distance zero) matches separately. We provide the necessary script to generate foreign keys for greater distances alongside the repository.


%% file: 40_analytics.tex
\section{Analytics}
\subsection{Summary Statistics}
We present a breakdown of the summary statistics of the schema present in the \textsc{SchemaDB}. These are presented as a series of truncated histograms due to very long tails. Our presentation of truncated tails provides ranged buckets at the tail of the axis to give an indication of the density and length of the tail. Particular interest should be paid to the recovered foreign keys with exact matching according to our na\"{i}ve heuristic, as keys were recovered in approximately $20\%$ of all samples lacking foreign keys, indicating that keys are frequently present but often unspecified.

\begin{figure*}[htb]
    \centering
    \includegraphics[width=\textwidth]{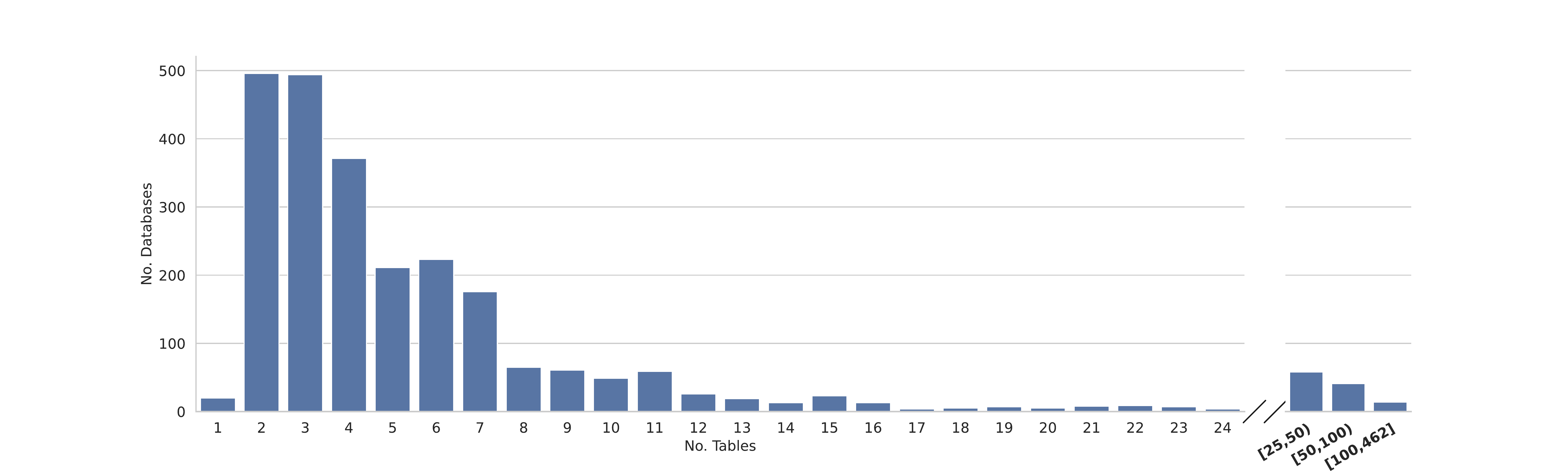}
    \caption{Database Size Statistics}
    \label{fig:db_size}
\end{figure*}

\begin{figure*}[htb]
    \centering
    \includegraphics[width=\textwidth]{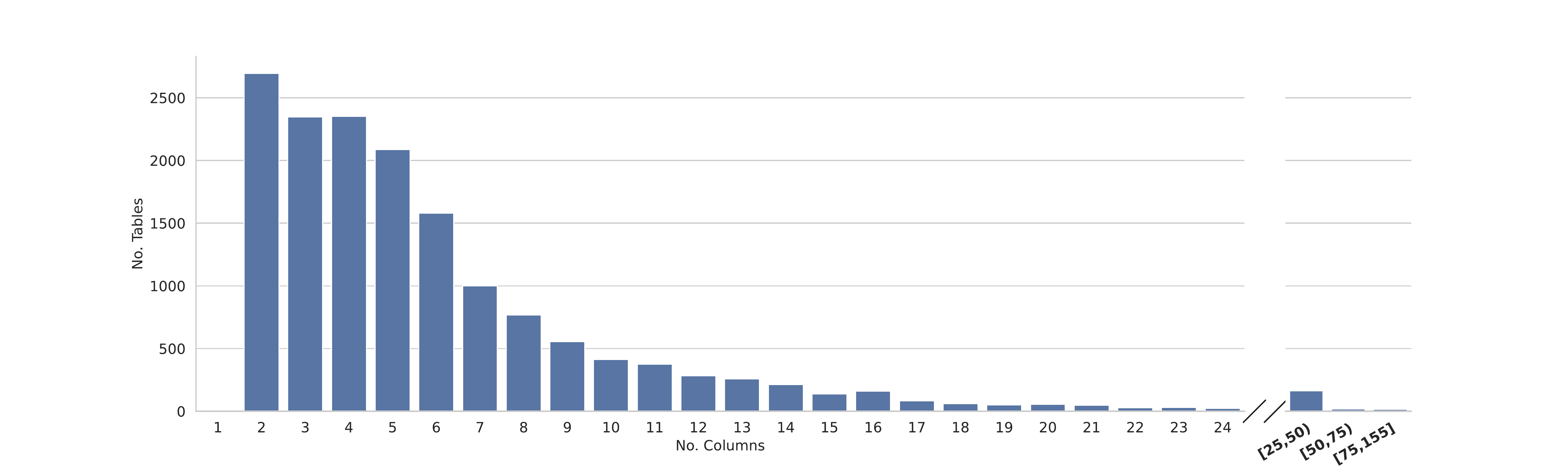}
    \caption{Table Size Statistics}
    \label{fig:tb_size}
\end{figure*}

\begin{figure*}[htb]
    \centering
    \includegraphics[width=\textwidth]{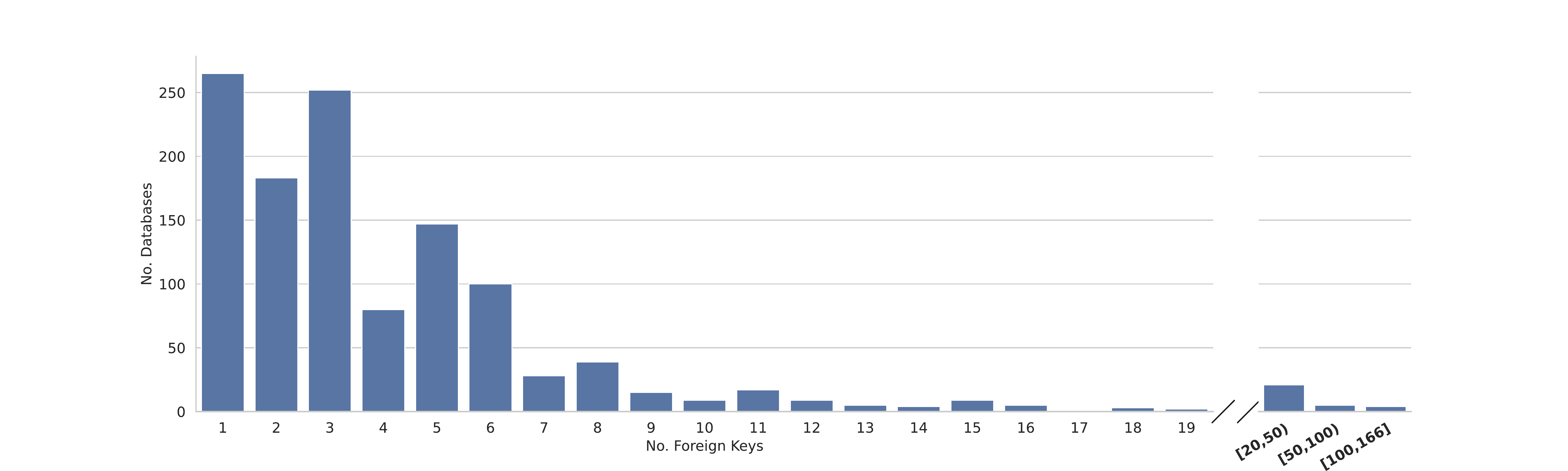}
    \caption{Foreign Key Count Statistics}
    \label{fig:fkct_size}
\end{figure*}

\begin{figure*}[htb]
    \centering
    \includegraphics[width=\textwidth]{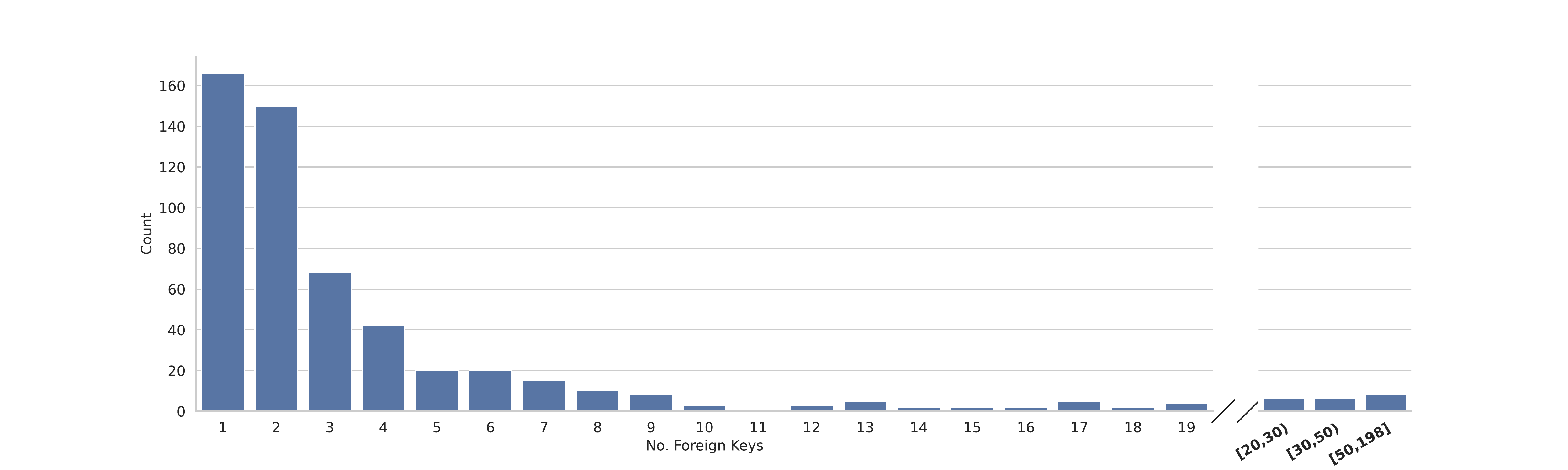}
    \caption{Recovered Foreign Key Count Statistics}
    \label{fig:rec_fkct_size}
\end{figure*}

In Figure~\ref{fig:db_size} we present a breakdown of the data in terms of the size of each database with respect to the number of tables present. In Figure~\ref{fig:tb_size} we further breakdown the size of tables with respect to the number of columns present. In Figure~\ref{fig:fkct_size} we show the prevalence of foreign keys as the number of these found in each database, excluding the zero cases. In Figure~\ref{fig:rec_fkct_size} we show, the number of recovered foreign keys recovered in roughly 500 of the cases where no foreign key was recorded.

%% file: 50_application.tex
\section{Research Potential \& Applications}
We present several possible downstream investigations enabled by the existence of this data set. To the best of our knowledge there are no existing analyses of relational schema in the wild. From a research perspective there are several interesting avenues worth pursuing. 

With respect to how databases are used in practice, we pose the following questions:
\begin{enumerate}
    \item How often are relational databases normalised (and to what degree) in practice?
    \item How often do specific entities appear and with what frequency do the co-occur? 
    \item By what names do common entities go by (i.e. person, user, customer, agent, employee) and can broad database purpose be determined from semantics?
\end{enumerate}
The first of these is of particular interest, as there is no way to determine which normal form a database is in without performing a manual analysis. A classifier that could determine the normal form of a schema would help in the development of optimally normalised databases. This would require labelling the schema for a supervised approach.

In multi-entity databases, some contextual signs of bad design are often evidenced by the following:
\begin{itemize}
    \item A single monolithic table\footnote{with regards to databases in relational engines. This convention is typically intentionally not followed in large flat-file highly distributed engines such a BigTable, Hadoop, etc.}
    \item Multiple entity types existing in the same table
    \item Absence of foreign keys linking clearly dependant columns across entities
\end{itemize}
We suggest that a database assistance agent that can identify normalisation status and detect entities could assist in the elimination of these types of issues.

In respect of the remaining questions (2 and 3) above, classification approaches in AI and ML could be considered. In particular the use of cutting edge language models such as GPT3~\cite{brown2020language} with fine-tuning on novel domains (e.g. Image-GPT~\cite{chen2020generative}) could assist with these determinations by way of clustering in the embedding space. 

This leads us to the question of automated generation, where there are a number of practical applications to consider. In particular:
\begin{enumerate}
    \item Is it possible to generate relational schema automatically from input text (such as a requirements specification)?
    \item Can existing data generation approaches be used in conjunction with schema generation to generate entirely novel databases with minimal to no prompting?
    \item Can this generation process be tuned for the creation of assets intended for cyber security, in particular cyber deception?
\end{enumerate}
Synthesising databases has application in cyber deception, where databases can be used as honeypots, or database elements as honeytokens~\cite{spitzner2003-1-honeypots,abay2019using}. Additionally, cyber research making use of realistic environments (such as reinforcement learning, or cyber ranges) could also benefit from the ability to use generated databases to increase (or decrease) the perceived realism of the environment. In terms of generative models, we also envision that the provided graph representations will enable approaches similar to that used in cutting edge arbitrary graph generation~\cite{you2018graphrnn, liao2019efficient, li2018learning, stier2021deepgg}.